\documentclass[sigconf]{acmart}
\usepackage{algorithm}
\usepackage{natbib}
\usepackage{graphicx}
\usepackage{algpseudocode}
\usepackage{amsmath}

\DeclareMathOperator*{\argmax}{argmax}

\newcommand*{\argmaxl}{\argmax\limits}

\AtBeginDocument{%
  \providecommand\BibTeX{{%
    \normalfont B\kern-0.5em{\scshape i\kern-0.25em b}\kern-0.8em\TeX}}}

\copyrightyear{2023} 
\acmYear{2023} 
\setcopyright{acmlicensed}\acmConference[WWW '23 Companion]{Companion Proceedings of the ACM Web Conference 2023}{April 30-May 4, 2023}{Austin, TX, USA}
\acmBooktitle{Companion Proceedings of the ACM Web Conference 2023 (WWW '23 Companion), April 30-May 4, 2023, Austin, TX, USA}
\acmPrice{15.00}
\acmDOI{10.1145/3543873.3584658}
\acmISBN{978-1-4503-9419-2/23/04}

\begin{document}

\title{CAViaR: Context Aware Video Recommendations}

\author{Khushhall Chandra Mahajan}
\authornote{Authors contributed equally to this research.}
\affiliation{%
  \institution{Meta Inc., USA}
  \streetaddress{}
  \city{}
  \state{}
  \country{}
  \postcode{}
}
\email{khushhall@meta.com}

\author{Aditya Palnitkar}
\authornotemark[1]
\affiliation{%
  \institution{Meta Inc., USA}
  \streetaddress{}
  \city{}
  \state{}
  \country{}
  \postcode{}
}
\email{aditpal@meta.com}

\author{Ameya Raul}
\authornotemark[1]
\affiliation{%
  \institution{Meta Inc., USA}
  \streetaddress{}
  \city{}
  \state{}
  \country{}
  \postcode{}
}
\email{araul@meta.com}

\author{Brad Schumitsch}
\affiliation{%
  \institution{Meta Inc., USA}
  \streetaddress{}
  \city{}
  \state{}
  \country{}
  \postcode{}
}
\email{bschumitsch@meta.com}

\renewcommand{\shortauthors}{Mahajan, Palnitkar, Raul, and Schumitsch, et al.}

\begin{abstract}
Many recommendation systems rely on point-wise models, which score items individually. However, point-wise models generating scores for a video are unable to account for other videos being recommended in a query. Due to this, diversity has to be introduced through the application of heuristic-based rules, which are not able to capture user preferences, or make balanced trade-offs in terms of diversity and item relevance. In this paper, we propose a novel method which introduces diversity by modeling the impact of low diversity on user’s engagement on individual items, thus being able to account for both diversity and relevance to adjust item scores. The proposed method is designed to be easily pluggable into existing large-scale recommender systems, while introducing minimal changes in the recommendations stack. Our models show significant improvements in offline metrics based on the normalized cross entropy loss compared to production point-wise models. Our approach also shows a substantial increase of 1.7\% in topline engagements coupled with a 1.5\% increase in daily active users in an A/B test with live traffic on Facebook Watch, which translates into an increase of millions in the number of daily active users for the product.

\end{abstract}



\maketitle

\section{Introduction}
The Facebook app is one of the largest platforms for discovering and watching videos online. Billions of users are able to find relevant videos from a corpus of videos of similar size, in the form of a personalized feed of videos, generated by sophisticated recommendation algorithms. 

Any recommender system has to strike a balance between serving relevant content that the user is most likely to enjoy, while also maintaining diversity in the entire slate of recommendations provided. 

Recommender models trained to predict user engagement tend to rank very similar videos at the top, given a large enough corpus of videos. However, presenting too much content that is similar to each other can lead to globally sub-optimal results at the session level, even though the user is likely to interact with each of the recommended videos when presented individually.

Like most recommender systems used in large-scale production systems, the video recommendation system at Facebook uses a deep neural network classification model to predict the likelihood of a user engaging on a particular video. Videos are ranked in descending order of predictions in the feed of videos presented to the user. In particular, the classification model computes the score for a tuple of user $u_i$ and a video $v_j$:

\begin{equation}
  s_{ij} = P(E(u_i, v_j) | F(u_i, v_j))
\end{equation}


where $E(u_i,v_j)$ denotes the event that the user $u_i$ positively interacts with the video $v_j$, and $F(u_i,v_j)$ $\in$ $R^\textsuperscript{n}$ is the n dimensional vector denoting features extracted for the user-video pair.
We employ a deep neural network based classification model to predict these probabilities for a user-video pair. This is a point-wise model, and only considers information regarding a video $v_j$ when computing the score for that video. This model does not incorporate information from other videos that will be served to the user above this video in the fully ranked feed of videos.
Due to this, the final list of videos can contain consecutive videos which are similar to each other, each of which individually have high predictions output by our classification model. However, we do not account for interactions between videos with each other, such that the predictions for a video $v_j$ should be lower than the computed score $s_{ij}$, after accounting for videos placed above that video in the feed.
We would thus like to actually compute the score 
\begin{equation}
  s'_{ij}=P(E(u_i, v_j) | F'(u_i, v_j, v_{(j-1)}, v_{(j-2)}, ..))
\end{equation}

where the videos $v_{(j-1)}$, $v_{(j-2)}$, ... are the videos placed successively above the video $v_j$ in a feed of videos. This new formulation of the score for ranking videos now accounts for all the videos the user will see in their feed, before encountering the video being currently considered. This score is expected to be more accurate compared to the original formulation $s_{ij}$ at predicting the occurrence of the event $E(u_i,v_j)$, and thus give us a better ranked feed of videos and a more engaging experience for the user. 

In this paper, we detail a technique that allows us to utilize this better formulation of a ranking score and adjust ranking accordingly to get closer to an optimal slate of recommendations.

This paper is structured as follows: We first present relevant work and key differences that differentiate our technique in section 2. Section 3 presents the system overview of our technique used to solve this problem, the Section 4 presents the experimental setup, and Section 5 presents the results obtained from using our technique, both through offline analysis and through online experimentation.

\section{Related work}
Majority of recommender systems are focused towards optimizing predictions for each item individually - i.e. predicting the probability of a user’s interactions with a given item. These point-wise estimations capture user interests effectively, and thus have been successfully leveraged to generate personalized rankings suited to each user’s tastes. It started with the use of Collaborative filtering \cite{Sarwar} and matrix factorization \cite{abcd} and continues with strong advancements with the use of deep neural networks \cite{Covington}. As mentioned in the introduction , the video recommendation system used in the Facebook app also leverages multiple deep neural networks where various signals from the user’s preferences and past history are combined with the video’s features to maximize the user’s engagement.
There has been a lot of work focused on incorporating novelty and diversity into recommendation systems \cite{Hijikata, McNee, Lathia, Vargas, EDBT, Ziegler}. Previous research has also been conducted to understand diversification in information retrieval  \cite{Clarke, SIGIR, Zhai, Rafiei, Santos}. We briefly summarize this work below.

\subsection{Novelty and Diversity}

Novelty relates to surfacing new experiences to users. For e.g. surfacing a football video for the first time to a user who generally likes basketball could be thought of a novel experience. There is previous work on utilizing user and feed context to show novel content to the user. This enables the recommender system to learn more about the user as well as enables the user to explore new content. 

Diversity relates to the differences between subsequent items in the current experience. For e.g. showing a mix of sports, news and entertainment videos in the feed yields a diverse experience. The primary motivation behind this kind of research is that by appropriately diversifying feed, one can improve the feed’s utility, thus maximizing the user’s satisfaction.
Initial work has focused on reducing redundancy through optimizing between relevance and similarity. For instance, Carbonell and Goldstein \cite{Carbonell} introduce the MMR (maximal marginal relevance) algorithm which involves iterating through each item at a time and scoring it based on a sum of the item’s relevance rating and a penalty for similarity with subsequent items. A common theme is to penalize items based on similarity using rules like in Ziegler et al \cite{Ziegler} or decaying similarity scores \cite{Perez}. Research in submodular functions also exists such as item selection based on submodular maximization in Tschiatschek et al \cite{Tschiatschek}. Teo et al \cite{Teo} use submodular diversity and item categories to re-rank items. 

\subsection{Generalized Contextual Ranking}

Our work adopts a different perspective - we utilize information informing the model of the context in which a video is placed when training a deep neural network. This enables us to personalize the treatment for each user, thus not only generalizing beyond the concepts of diversity and novelty, but also allowing for personalized settings of such dimensions to suit user interests. There has been similar work to exploit a personalized notion of Diversity where Mark et al \cite{Wilhelm} experimented with a DPP based approach that incorporates pointwise and similarity scores on a large scale recommendation system like Youtube. 
Our work distinguishes itself by being reliant on point-wise classification models for the introduction of diversity. This allows us to train and serve models in production recommendation systems without any costly changes to the infrastructure or tooling. 
Our work is also unique in the sense that diversity is not treated as an objective independent of relevance, or user engagement. We see that addition of diversity is a byproduct of using models that are capable of improving relevance through the use of features that capture diversity-related information.

\section{System Design}

When a user visits a video recommendation surface in the Facebook app, we initially generate a list of hundreds of video candidates that the user might be interested in. This list of videos is passed through computationally intensive deep learnt models which predict the score $s_{ij}$ for a user-video pair. This stage of serving recommending videos is referred to as the main ranking pass, owing to its computational complexity.
After this main ranking pass, we introduce a contextual pass, which allows us to compute $s’_{ij}$ and re-rank videos based on the updated scores accounting for ‘contextual information’, i.e. information derived from videos preceding the video $v_j$ in the feed. To compute and utilize $s’_{ij}$ in a computationally feasible way, we employ a greedy approach described in the following algorithm.
Assume that our main ranker has generated a list of K videos, which we are to re-rank for a user u. $v_j$ represents a video at position $j$ after the main ranking pass, while $v'_j$ represents the video at position $j$ after the re-ranking pass.

\begin{algorithm}
\caption{Re-ranking a feed using a contextual model}\label{alg:cap}
\begin{algorithmic}
\For{$i \in 1$:$K$}
\For{$j \in $i:$i + $w}
\State $s’_{uv_j} \gets P(E(u, v_j) | F'(u, v_j, v_{(j-1)}, v_{(j-2)}, ..))$
\EndFor
\State $v'_j \gets \argmaxl_{x} \| s'_{uv_x}  \|$
\EndFor
\end{algorithmic}
\end{algorithm}

This algorithm allows us so slot videos in each successive position with the highest score $s’_{ij}$, given the characteristics of the videos slotted above that position. 

\begin{figure}[h]
  \centering
  \includegraphics[width=\linewidth]{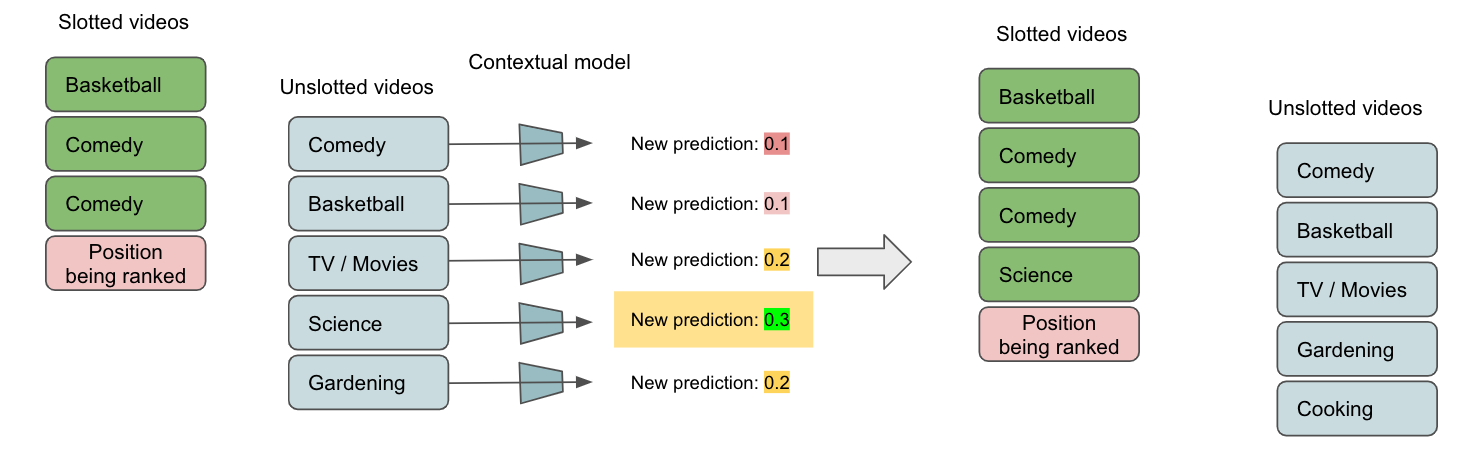}
  \caption{After the main ranking pass, we successively slot videos in the final ranked order.}
  \Description{After the main ranking pass, we successively slot videos in the final ranked order.}
\end{figure}

This approach allows us to use a prediction model very similar to the one used in the main ranking pass, but employ it to re-rank the feed of videos to introduce diversity. However, this method has the potential of introducing latency regressions, due to the complexity O(K * n), where K is the number of positions, and n is the search window.

To avoid incurring this latency, we propose a demand based re-ranking approach to mitigate this latency concern. Although our ranking system yields K videos to the user, we only show a much smaller subset p of them to the user (p << K). This is because users often have limited real-estate on their devices and it is impossible to show all K items to the user. Moreover users also spend time focusing on the top p videos. 
Our technique utilizes this user behavior to reduce perceived latency. Rather than applying re-ranking on all K videos in a single go, we apply it for only the first p videos. As the user consumes content and scrolls through their feed, we trigger subsequent iterations of the contextual pass on the p videos at a time. This enables us to reduce the latency impact while being able to re-rank feed using classification based models.

\subsection{Contextual Features}
Contextual features are defined as the features based on videos surrounding a given video in a ranked feed. Assume a point-wise model which ranks the videos in an order based on some score. Given this list of ordered videos, we design features to capture the contextual information. Some examples of such features:

1. Averaged embeddings: We use pretrained video embeddings for each video. These embeddings could be designed for video understanding embeddings, etc. We extract the average of the embeddings for every video in a window of size k. 

2. Similarity features: We take the embedding of the video into consideration (say at position i). We then compute the dot product of the current video with the k videos above it, and consider each dot product to be a similarity score. Now we can have two separate features, in the form of the average of the k similarity scores, and each score extracted separately.

3. Video Topic: We often have topics tagged on videos through automated classifiers. We can extract information on the topic overlap betwween a given video and the videos surrounding that video, and use it as a measure of diversity.

In all these feature designs, there is a common philosophy. The method to capture the context should be computationally inexpensive and fit well within the framework of such large models. Each of the above features uses the existing framework and relies on embeddings which are already heavily used in large-scale models.

\subsection{Contextual Model}

Here, we describe the model used to compute the scores s'(ij) used for re-ranking. Since this model utilizes contextual features described above to augment its predictions, we call this a contextual model.
We used a deep neural network based model which comprises of user side features and video side features. The model uses an embedding layer to convert sparse features into values. The model architecture comprises of multiple dense layers using ReLU as an activation. The final layer has multiple objectives, each of which maps to a positive user engagement event. We finally use a cross entropy loss. Besides contextual features, we include more user and video based features as inputs to the model. 


\section{Experiment}
We trained two different point wise models. The first, a baseline model, is without any changes and the second, contextual model includes the contextual features. We added ten additional contextual features in our contextual models. 
Each of these models uses Adam optimizer with an initial learning rate of 0.005 and batch size of 128. We train our model on the 21 days of data initially and then train it recurrently on each day of additional data. We do a single training pass over our data.

\section{Result}

\subsection{Model Calibration}

Model calibration is a metric we commonly use to evaluate models and understand if they are biased towards over-predicting for some videos over other videos. Calibration is defined as:

\begin{equation}
  calibration=\frac{\sum{prediction}}{\sum{label}}
\end{equation}

A desirable property for a prediction model is to be well calibrated, which is defined as having a calibration equal to 1 over all subsets of our data. 
If a model is not calibrated for a subset of videos or users which have a particular property, the model is either over or under predicting for those items. In such a case, model calibration can be fixed, and performance improved by adding this property as an input feature to the model.

To understand if diversity is a problem in our recommendation system, we plot model calibration against features capturing the diversity of feed using different contextual features. We use the most important user engagement event, a binary classification event, to derive calibration.

We choose a similarity score as as measure of diversity to plot the calibration against.
We use pre-trained embeddings assigned to each video, which denote similarity in content and topics. The similarity score is computed by taking the dot product of a video’s embedding with the average embedding of the previous 5 videos in feed. This score is bucketized, and calibrations plotted against these buckets. 

In Figure [2], the first graph shows a clear trend- higher the similarity, the more over-calibrated our predictions. We also see that there's a large scope for improvement in a significant percent of our data: For example, in the case of first graph, we see 40\% of our samples with similarity bucket > 30 have a mis-calibration of more than 4\%. We should see significant gains from fixing this calibrations.

To confirm this hypothesis, we check the calibration of the contextual model trained using contextual features. We expect that this model should be able to be well calibrated across all values of similarity scores.
The second graph in Figure [2] shows that indeed the calibration of the model is now fairly constant when plotted against the similarity score, aligned with our hypothesis. 

\begin{figure}[h]
  \centering
  \includegraphics[width=\linewidth]{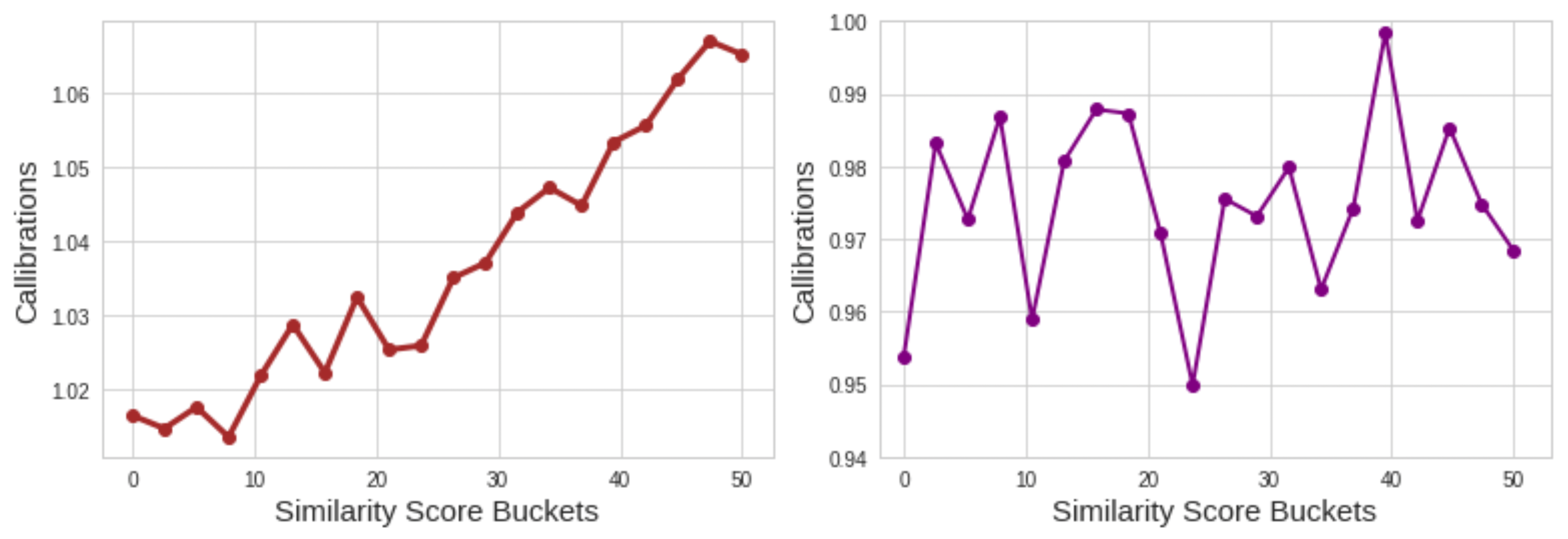}
  \caption{Calibrations for a user engagement event from the main ranking pass against measures of diversity (left), and from the contextual model (right) }
  \Description{Calibrations for a user engagement event from the main ranking pass against measures of diversity (left), and from the contextual model (right)}
\end{figure}

\subsection{Offline performance}

Next, we would like to see if improving the model calibration leads to an improvement in other model evaluation metrics.
We use normalized entropy to measure the model’s offline performance on the binary prediction task. Normalized Entropy (NE) is defined as the predictive log-loss per impression, divided by the entropy of the background CTR (click-through rate). The background CTR is the average empirical CTR of the training data and lower normalized cross-entropy is better.

    \begin{figure}[h]
  \centering
  \includegraphics[width=0.8\linewidth]{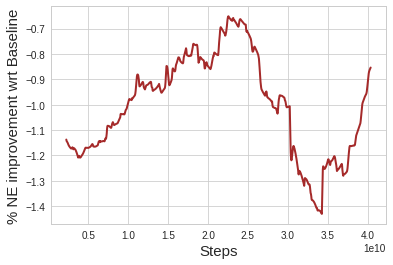}
  \caption{Progression of percentage improvement in offline normalized entropy for the contextual model as training progresses, using the main ranking model as a baseline}
  \Description{Percentage improvement in the NE for one of the engagement model compared to baseline}
\end{figure}

When comparing the model performance of a model utilizing contextual features against the baseline model trained without these features, we see a NE improvement on three different engagement prediction models. The improvement over baseline models are 1.2\%, 0.85\% (as seen in Figure [3]), and 1.4\%.  Offline gains are a strong indicator that our model will show online gains in the A/B test.

\subsection{A/B testing}
When evaluated in an online A/B test, we see that the contextual model leads to significant improvements. In particular we observe a 1.7\% improvement in user topline engagement metrics, as seen in Figure [4]. We also see an increase of 1.6\% in daily activity, which is a metric measuring distinct users engaging on videos in a day. This accounts for a significant increase, given the baseline of billions of user video engagements per day on the Facebook app. Furthermore, we see an increasing trend in the metrics when measured on a daily basis, suggesting that users show increasingly accruing satisfaction with the recommendations. 

\begin{figure}[h]
  \centering
  \includegraphics[width=\linewidth]{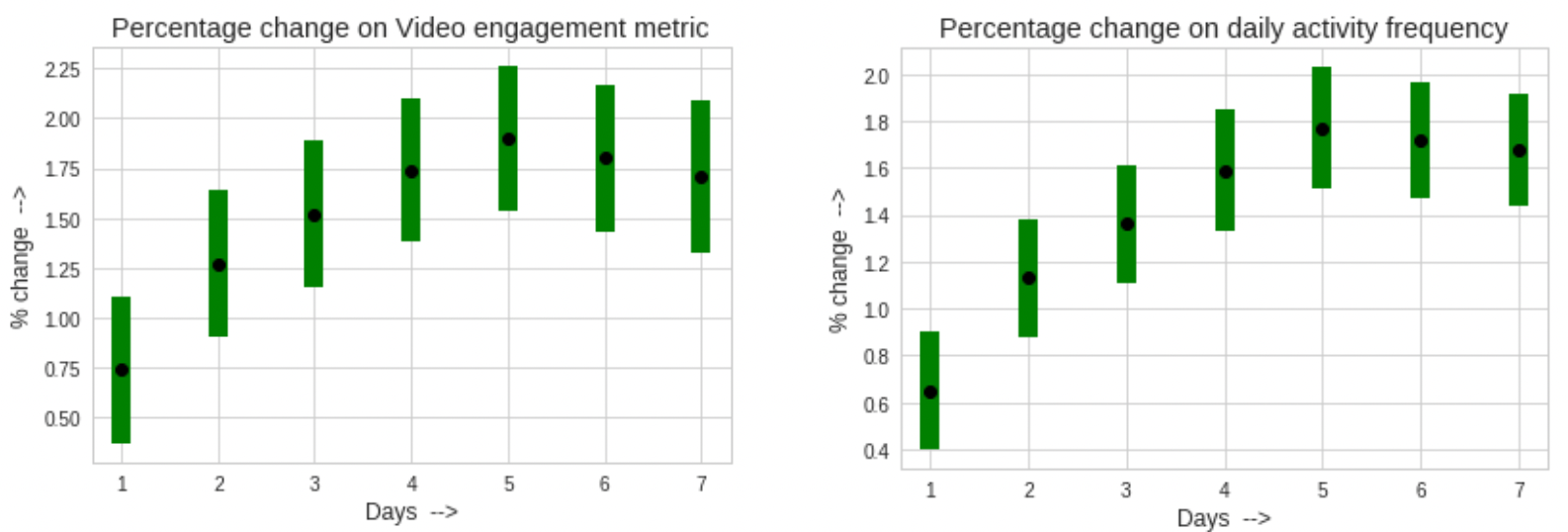}
  \caption{Improvement in engagement metrics in an online A/B test.}
  \Description{Improvement in engagement metrics in an online A/B test.}
\end{figure}

\section{Conclusion}
Diversification of items is a persistent challenge for any recommendation system. Most methods of introducing diversity in recommendation systems do so through the use of heuristics, or treat diversity as an objective that is at odds with user engagement. 
In this paper, we prove that optimizing for user engagement can also introduce diversity, as long as we make our models aware of diversity-related features. By not treating diversity as an objective separate from user engagement, we do not have to encode arbitrary trade-offs amongst diversity and relevance in our system. The models are able to introduce diversity only for users and items that would be negatively impacted due to lack of it.
Furthermore, our method is designed such that it can be introduced to any large scale recommender system using point-wise models similar to those currently being used in the recommendation stack. This gives us the ability to use all the tools and supporting infrastructure used to serve such models without any significant changes.

\section{Next steps}
We would like to integrate the contextual model with our main ranking pass model, by co-training the two models. In the final output layer, the combined model can output both a non-contextual prediction, and a contextual prediction given an additional set of contextual features. This way, we can pre-compute the non-contextual parts of the model and cache them to be used later in the contextual pass to reduce latency and CPU costs.

Another way to improve this approach would be to encode sequential information when extracting contextual features, through the use of sequential models like LSTMs to generate embeddings, rather than use averages. Thus, using point-wise models to improve diversity and relevance is an ongoing vector for improvements to the user experience on Facebook Watch.

\bibliographystyle{main}
\bibliography{main}

\end{document}